\documentclass[sigconf, nonacm]{acmart}

\usepackage{pvldb}

\renewcommand\vldbdoi{XX.XX/XXX.XX}
\renewcommand\vldbpages{XXX-XXX}
\renewcommand\vldbavailabilityurl{https://github.com/RazeenLI/JevNexus.git} 

\begin{document}
\title{Correspondences as Decisions: JevNexus for Decision-Centric Schema Matching}

\author{Runze~Li}
\orcid{0009-0004-1695-1422}
\affiliation{%
  \institution{University of New South Wales}
  \city{Sydney}
  \country{Australia}
}
\email{runze.li1@unsw.edu.au}

\author{Hanchen~Wang}
\orcid{0000-0003-3158-9586}
\affiliation{%
  \institution{University of Technology Sydney}
  \city{Sydney}
  \country{Australia}
}
\email{hanchen.wang@uts.edu.au}

\author{Ying~Zhang}
\orcid{0000-0002-2674-1638}
\affiliation{%
  \institution{University of Technology Sydney}
  \city{Sydney}
  \country{Australia}
}
\email{ying.zhang@uts.edu.au}

\author{Wenjie~Zhang}
\orcid{0000-0001-6572-2600}
\affiliation{%
  \institution{University of New South Wales}
  \city{Sydney}
  \country{Australia}
}
\email{wenjie.zhang@unsw.edu.au}

\begin{abstract}
Schema matching increasingly uses generative language models to rerank retrieved
column candidates, although the underlying task is a bounded correspondence
decision. We present \textsc{JevNexus}, which combines typed pairwise decisions with
schema/instance evidence and invokes listwise refinement only when the evidence
disagrees and the fused margin is small. The evaluation covers 561 cases from
six benchmark families. \textsc{JevNexus} obtains dataset-macro MRR and Hits@1 of
0.930 and 0.909, compared with 0.926 and 0.903 for \textsc{Magneto}, while
reducing mean latency from 123.452 to 15.929 seconds (7.750$\times$). Paired
analysis finds no statistically significant difference in either MRR or
Hits@1. The
gate invokes listwise refinement for only 5.665\% of source columns and avoids
the degradation caused by unconditional refinement. Code and experimental
artifacts are available at \url{https://github.com/RazeenLI/JevNexus}.
\end{abstract}

\maketitle



\section{Introduction}
\label{sec:introduction}

The continuing growth of structured data in scientific repositories, enterprise platforms, and public data ecosystems creates substantial opportunities for cross-source analytics, scientific discovery, and data-driven decision making~\cite{li2023proteogenomic,fernandez2018aurum,cafarella2008webtables,deng2024lakebench,li2025fadti}.
Realizing these opportunities, however, often requires independently developed datasets to be integrated before they can be queried or analyzed together.
For example, cancer consortia harmonize genomic, proteomic, and clinical data to support pan-cancer studies~\cite{li2023proteogenomic}; enterprise discovery systems expose relationships among tables distributed across organizational data repositories~\cite{fernandez2018aurum}; and Web- and data-lake-scale systems identify tables that can be joined or unioned for downstream analysis~\cite{cafarella2008webtables,deng2024lakebench}.
These workflows depend on schema matching, the task of identifying attributes in heterogeneous schemas that represent the same underlying concepts~\cite{rahm2001survey,madhavan2001cupid,do2002coma}.

\begin{figure}[t]
    \centering
    \includegraphics[width=\columnwidth]{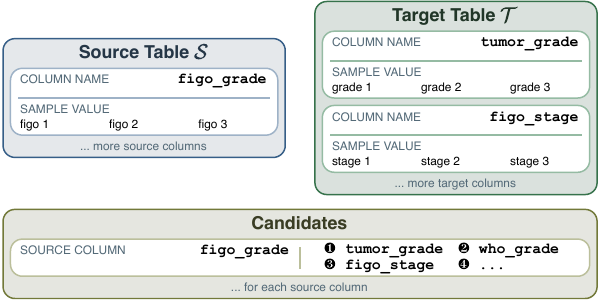}
    \caption{A schema-matching example with conflicting evidence.}
    \label{fig:motivating-example}
\end{figure}

Formally, given a source schema $S$ and a target schema $T$, schema matching seeks a set of valid correspondences $M \subseteq S \times T$ and, in ranking-oriented settings, places the valid targets for each source attribute early in the resulting order.
When $T$ is large, a localization stage can first restrict the search to a bounded candidate set $C(s) \subseteq T$ for each $s \in S$.
Once these candidates have been identified, the fundamental operation is a collection of constrained questions: for each $t \in C(s)$, do $s$ and $t$ represent the same underlying schema attribute?
This operation is naturally expressed as a bounded correspondence decision, with each pair assessed on its own merits and multiple or no candidates allowed to match, rather than as an open-ended generation task.
Recent LLM-based matchers instead approximate the operation by asking a generative model to emit candidate names together with textual or JSON-encoded similarity scores~\cite{liu2025magneto}.
Although such generation can produce effective rankings, it represents the underlying judgments indirectly through autoregressive decoding, output parsing, and generated numeric values.
We therefore investigate a decision-centric formulation in which localized candidates are primarily evaluated through separately scored, non-exclusive correspondence decisions.

Figure~\ref{fig:motivating-example} illustrates the distinction using a source column named \texttt{figo\_grade} and two plausible target columns, \texttt{tumor\_grade} and \texttt{figo\_stage}, following the type of ambiguity found in the GDC schema-matching setting~\cite{liu2025magneto}.
The token \texttt{figo} provides evidence for \texttt{figo\_stage}, whereas the attribute type expressed by \texttt{grade} provides evidence for \texttt{tumor\_grade}, and the sampled values may not fully eliminate this ambiguity.
The correspondence therefore cannot be resolved reliably from a single surface-similarity signal.
Evaluating the two pairs separately makes the plausibility of each correspondence explicit, while comparing the candidates remains useful when their evidence is closely balanced.
The example separates three operations that a matching system must coordinate: judging individual correspondences, integrating complementary evidence, and comparing unresolved candidates.

Classical schema matchers combine lexical, structural, and instance-level signals, reflecting the long-standing observation that no single form of evidence is reliable across all schema pairs~\cite{rahm2001survey,madhavan2001cupid,do2002coma,melnik2002similarity}.
Although these methods preserve complementary evidence, manually designed similarities often struggle with correspondences that require broader semantic understanding.
Pretrained language models extend this semantic coverage by learning representations or matching functions for schema elements~\cite{zhang2023plm,tu2023unicorn,du2024isresmat}.
More recently, retrieve--rerank systems such as \textsc{Magneto} use an efficient semantic model to localize candidates and a generative LLM to produce their final scores or ordering~\cite{liu2025magneto}.
This progression improves semantic matching and makes LLM reasoning feasible over wide schemas, but it leaves three gaps.
First, the final correspondence judgment is expressed indirectly through generated output rather than through an explicit matching decision.
Second, the move toward a single semantic reranker does not explicitly preserve the complementary roles of semantic, schema-level, and instance-level evidence.
Third, uniformly reranking every localized candidate set does not distinguish cases that require joint comparison from those for which the available evidence already supports a clear ordering.

These gaps lead directly to three challenges for decision-centric schema matching.
The first is to represent the plausibility of every localized pair directly, without autoregressive score generation or normalization across candidates.
The second is to integrate semantic judgments with complementary schema and instance evidence when different signals provide different views of a candidate, as in Figure~\ref{fig:motivating-example}.
The third is to retain joint candidate comparison for genuine residual ambiguities without allowing a listwise model to reconsider every otherwise confident ranking.
Meeting these challenges requires pairwise judgment, evidence integration, and cross-candidate comparison to remain distinct but coordinated operations.

We introduce \textsc{JevNexus}, a progressive schema-matching architecture built around this separation.
The pipeline begins by localizing a bounded candidate set for each source attribute, separating candidate recall from final correspondence judgment.
A typed decision operator assigns every localized pair a separately scored, non-exclusive probability that the two columns represent the same underlying attribute.
These semantic decisions are combined with schema- and instance-level scores so that the default ranking reflects complementary evidence rather than a single model output.
When the evidence sources prefer different candidates and the highest fused scores remain close, an evidence-aware gate sends only the leading alternatives to a compact listwise reranker.
Otherwise, the fused ranking bypasses listwise refinement and remains unchanged.
In this way, \textsc{JevNexus} uses pairwise decisions to establish correspondence plausibility, heterogeneous evidence to form the default order, and joint reasoning only to resolve remaining local ambiguities.

We evaluate \textsc{JevNexus} on 561 schema-matching cases from GDC-SM and five Valentine benchmark families, covering 13,522 source columns and 8,944 ground-truth correspondences.
Across the six benchmark families, \textsc{JevNexus} achieves the highest dataset-macro MRR and Hits@1 among the evaluated methods and is statistically comparable to the strongest generative method over all 561 cases.
It obtains the highest or tied-highest MRR on GDC, Magellan, and WikiData and remains competitive on the other benchmark families.
More importantly, \textsc{JevNexus} reduces mean online runtime from 123.452 to 15.929 seconds per case, yielding a $7.750\times$ speedup over generative reranking.
Ablations show that complementary schema and instance evidence provides the largest effectiveness gain, while unconditional listwise refinement is less effective than the proposed selective strategy.
The evidence-aware gate produces the strongest aggregate \textsc{JevNexus} variant while activating for only 5.665\% of source columns, showing that most localized decisions do not require joint reranking.

We make the following contributions:
\begin{itemize}
    \item \textbf{Decision-centric formulation.}
    We formulate localized schema matching as explicit, separately scored, and non-exclusive correspondence decisions rather than autoregressive similarity-score generation.

    \item \textbf{Evidence-aware progressive architecture.}
    We develop \textsc{JevNexus}, which combines typed semantic decisions with complementary schema and instance evidence and invokes listwise comparison only under simultaneous evidence disagreement and fused-ranking uncertainty.

    \item \textbf{Controlled full-benchmark evaluation.}
    We compare \textsc{JevNexus} with traditional, pretrained-neural, and generative schema matchers across 561 cases and analyze ranking effectiveness, efficiency, component contributions, routing behavior, dataset heterogeneity, and remaining failure modes.
\end{itemize}

\begin{figure*}[t]
    \centering
    \includegraphics[width=\linewidth]{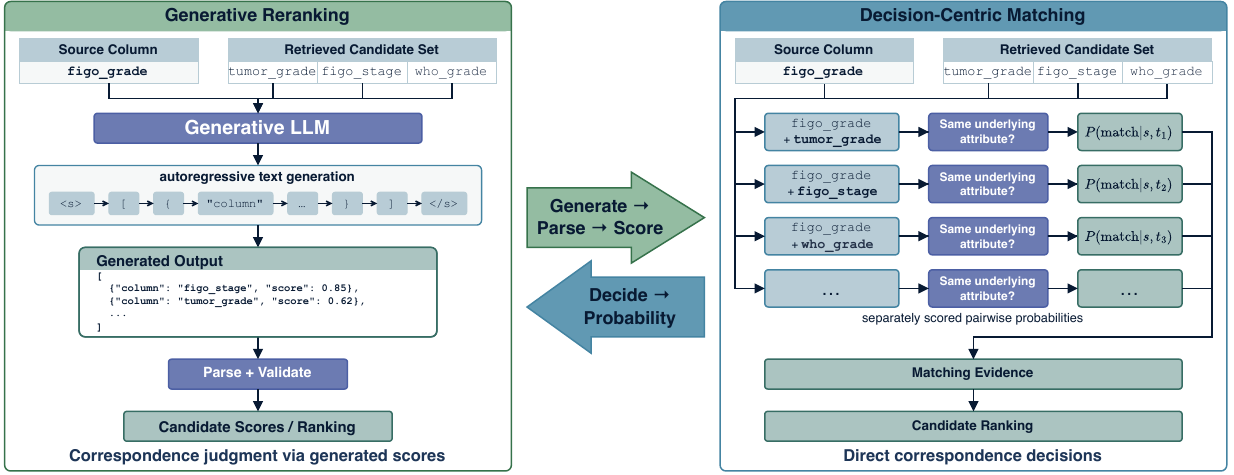}
    \caption{Generative and decision-centric inference over the same localized candidates.}
    \label{fig:compare}
\end{figure*}

\section{Problem Formulation and Motivation}
\label{sec:problem}

\subsection{Schema Matching and Ranked Output}

Let a source table $A$ have schema $S=\{s_1,\ldots,s_m\}$ and a target table $B$ have schema $T=\{t_1,\ldots,t_n\}$.
For a column $c\in S\cup T$, let $N(c)$ denote its name and $V(c)$ the available sample of observed non-null values, which may be empty.
We write the observable column evidence as
\[
\phi(c)=\bigl(N(c),V(c)\bigr).
\]
A schema correspondence is a pair $(s,t)\in S\times T$ whose columns represent the same underlying attribute according to the benchmark semantics~\cite{rahm2001survey,liu2025magneto}.
The ground-truth correspondence relation is denoted by $G\subseteq S\times T$, and the valid targets for a source column $s$ are
\[
G(s)=\{t\in T\mid(s,t)\in G\}.
\]
This relation does not impose a one-to-one constraint: depending on the matching setting, $G(s)$ may contain zero, one, or multiple target columns.

Following ranking-oriented schema matching~\cite{koutras2021valentine,liu2025magneto}, a matcher produces for every source column $s\in S$ a complete ordering
\[
\pi_s=\langle t_{(1)},t_{(2)},\ldots,t_{(n)}\rangle
\]
of the target columns.
The ranking may later be thresholded or transformed into a discrete mapping, but our primary objective is to place the members of $G(s)$ as early as possible in $\pi_s$.
This ranking formulation separates the quality of correspondence evidence from any application-specific threshold used to accept or reject matches.

\subsection{Candidate-Localized Matching}

When the target schema is wide, applying an expensive semantic operator to every pair in $S\times T$ is unnecessary.
A localization operator $\mathcal{L}$ therefore returns at most $k$ candidate targets for each source column:
\[
C_k(s)=\mathcal{L}(s,T),
\qquad
|C_k(s)|\leq k.
\]
Localization and correspondence judgment nevertheless solve different problems: $\mathcal{L}$ identifies a high-recall neighborhood, whereas the subsequent matcher must discriminate among the candidates in that neighborhood.
If $G(s)\cap C_k(s)=\varnothing$, no downstream decision or reranking operator restricted to $C_k(s)$ can recover a valid target at the head of the ranking.
Candidate localization therefore determines a recall ceiling, while the quality of the ordering within $C_k(s)$ determines how effectively the available correspondences are surfaced.

\subsection{Generated versus Decision-Based Evidence}

Once $C_k(s)$ has been obtained, generative reranking and decision-centric matching receive the same source column, candidate columns, and observable column evidence, and both ultimately support an ordering over the candidates.
They differ in how the correspondence evidence underlying that ordering is represented and obtained, as illustrated in Figure~\ref{fig:compare}.
In generative reranking, a model jointly inspects the source and candidate set and autoregressively emits candidate identifiers together with textual or JSON-encoded scores.
The matching system then parses and validates the generated artifact before converting it into candidate scores and a ranking.
Although this procedure can produce effective rankings, its correspondence judgments are represented indirectly through generated tokens.

A decision-centric formulation instead poses the bounded matching question directly for each localized pair: do $s$ and $t$ represent the same underlying attribute?
The resulting answer is expressed as bounded pairwise evidence and can be consumed without generating or parsing a textual score.
These direct judgments provide semantic evidence for constructing the candidate ranking and, when useful, for combination with other matching signals.

\begin{figure*}[t]
    \centering
    \includegraphics[width=\linewidth]{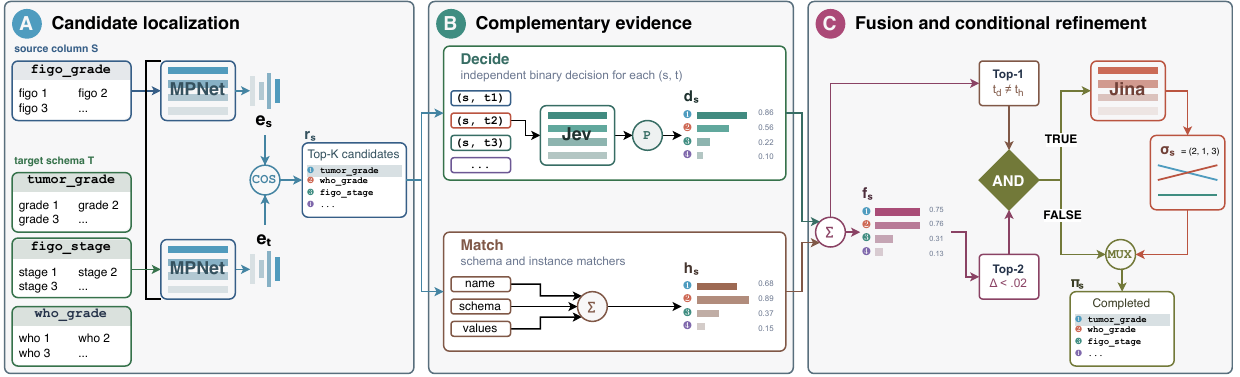}
    \caption{Overview of the \textsc{JevNexus} matching pipeline.}
    \label{fig:framework}
\end{figure*}

\subsection{Decision-Centric Matching Objective}

For each localized pair $(s,t)$ with $t\in C_k(s)$, define the correspondence label
\[
y_{s,t}=\mathbf{1}\bigl[(s,t)\in G\bigr].
\]
A decision-centric matcher seeks bounded semantic evidence
\[
q(s,t)\in[0,1]
\]
whose magnitude reflects the plausibility that $y_{s,t}=1$ given $\phi(s)$ and $\phi(t)$.
Each $q(s,t)$ is obtained as a separately scored correspondence judgment rather than as one component of a multiclass distribution over $C_k(s)$.
Consequently, the values $\{q(s,t):t\in C_k(s)\}$ are not constrained to sum to one: multiple candidates may receive strong correspondence evidence, and all candidates may receive weak evidence.

The values $q(s,t)$ represent semantic decision evidence rather than prescribing the complete matching algorithm or its final score.
A matcher may combine them with complementary schema- or instance-level evidence before producing $\pi_s$.
The resulting problem is to rank valid targets early while bounding expensive inference to the localized candidate sets.
Section~\ref{sec:method} presents \textsc{JevNexus}, which realizes this formulation through typed correspondence decisions, complementary evidence integration, and selectively invoked cross-candidate refinement.

\section{JevNexus}
\label{sec:method}

\subsection{Architecture Overview}

\textsc{JevNexus} is a progressive schema-matching architecture that separates candidate localization, direct correspondence judgment, complementary evidence integration, and residual cross-candidate comparison.
For each source column $s$, the pipeline first localizes a bounded candidate neighborhood $C_k(s)$ and then assigns every localized pair two functionally distinct signals: typed semantic decision evidence $d(s,t)$ and complementary schema/instance evidence $h(s,t)$.
The two signals are integrated into a default score $f(s,t)$, and a routing criterion determines whether the fused head is sufficiently ambiguous to justify local listwise comparison.
The resulting localized order is finally combined with the untouched target-schema tail to produce the complete ranking $\pi_s$.

Figure~\ref{fig:framework} summarizes this information flow.
Localization is responsible for candidate recall, typed decisions express pairwise semantic correspondence, complementary evidence captures lexical and value regularities, and listwise refinement is restricted to residual local ambiguities.
This decomposition makes the contribution of each evidence source and the effect of selective routing directly observable.

\subsection{Candidate Localization}
\label{sec:localization}

Candidate localization instantiates the operator $\mathcal{L}$ introduced in Section~\ref{sec:problem}.
Each column is represented using its normalized name and a deterministic sample of observed non-null values.
The representation is serialized into a compact text sequence and encoded in a shared semantic space~\cite{reimers2019sentencebert,song2020mpnet}.
For a source column $s$ and target column $t$, the retrieval score is cosine similarity,
\[
r(s,t)=\cos\bigl(e(s),e(t)\bigr),
\]
where $e(\cdot)$ denotes the resulting column embedding.

The localizer supplements semantic retrieval with exact normalized-name matches, removes pairs below the configured similarity threshold, orders the surviving targets by retrieval evidence, and retains at most $k$ candidates.
Exact-name evidence prevents simple but important correspondences from being lost because of representation noise.
All ordering operations use deterministic tie breaking.

The retrieval score $r(s,t)$ determines membership and initial order in $C_k(s)$ but is not treated as final correspondence evidence by the subsequent decision and fusion stages.
The localizer also retains the target columns outside $C_k(s)$ in target-schema order so that the pipeline can later construct a complete ranking.

\subsection{Typed Correspondence Decisions}
\label{sec:decision}

The typed decision operator realizes the direct correspondence semantics established in Section~\ref{sec:problem}.
Rather than requesting a serialized list of candidate scores, it presents each localized source--target pair as a bounded binary question whose positive outcome denotes that the two columns represent the same underlying attribute.

\subsubsection{State--Question Construction}

For a source column $s$, \textsc{JevNexus} constructs a source state
\[
x_s=\operatorname{serialize}\bigl(N(s),V(s)\bigr).
\]
For every $t\in C_k(s)$, it constructs a candidate-specific typed question
\[
Q_{s,t}=\operatorname{question}\bigl(N(t),V(t)\bigr)
\]
that asks whether the candidate represents the same underlying attribute as the source.
The positive criterion permits different surface forms when the real-world attribute is equivalent, whereas the negative criterion covers distinct attributes despite lexical or value-level similarity.

\begin{figure}[t]
\includegraphics[width=\columnwidth]{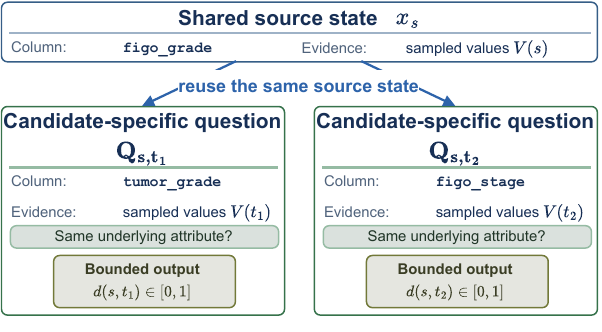}
    \caption{Typed correspondence decisions for two localized candidates.}
    \label{fig:state-question}
\end{figure}

Figure~\ref{fig:state-question} instantiates this interface with the running example from Figure~\ref{fig:motivating-example}.
The source profile is shared, but each candidate supplies a separate question and receives its own bounded match value.
The state and questions contain only observable column evidence: they exclude retrieval scores and ranks, dataset identifiers, benchmark metadata, and ground truth.
Candidate identifiers serve only to associate returned values with questions and carry no semantic or ranking information.

All questions for a source column can be submitted in one batched decision request while retaining candidate-specific contexts.
The source state is therefore encoded once, and each question exposes only the evidence of its own candidate rather than the complete candidate list.
This organization preserves separately scored pairwise semantics without issuing one external request per pair.

\subsubsection{Bounded Decision Inference}

For every localized pair, the decision operator returns
\[
d(s,t)
=
P_{\theta}\!\left(y_{s,t}=1\mid x_s,Q_{s,t}\right)
\in[0,1].
\]
The value is obtained directly from the typed binary decision interface rather than generated as a numeric string, candidate list, or explanation.
Consequently, the matching pipeline does not need autoregressive answer decoding or score extraction from free-form text.

Candidates are evaluated as separate questions and are not normalized with a softmax over $C_k(s)$.
The values $\{d(s,t):t\in C_k(s)\}$ are therefore not constrained to sum to one: multiple candidates may receive strong evidence, and all candidates may receive weak evidence.

The output contract requires exactly one finite value in $[0,1]$ for every submitted candidate identifier.
Missing identifiers, unexpected identifiers, duplicate fields, and out-of-range values are treated as invalid outputs rather than silently converted into correspondence scores.

\subsection{Complementary Schema and Instance Evidence}
\label{sec:complementary}

Direct semantic decisions do not exhaust the evidence available for schema matching.
Column names can exhibit useful lexical regularities, and observed values can reveal correspondences that remain ambiguous from language-level semantics alone.
\textsc{JevNexus} therefore computes a second score $h(s,t)$ with a non-generative matching operator that combines schema- and instance-level signals, following the classical principle of composing heterogeneous match functions~\cite{do2002coma,engmann2007comapp}.

\subsubsection{Schema-Level Evidence}

For each source--target pair, the schema-level branch compares normalized column names with character-trigram and token-based similarities.
The two measures are complementary: trigram overlap captures local spelling variation, while token similarity is more robust to separators, reordered fragments, and multiword names.
Their maximum forms the name score
\[
h^{\mathrm{name}}(s,t)
=
\max\!\left\{
h^{\mathrm{tri}}(s,t),
h^{\mathrm{tok}}(s,t)
\right\}.
\]
Because the evaluated inputs are flat relational tables, this branch uses column-level name evidence rather than introducing artificial hierarchy or neighborhood structure.

\subsubsection{Instance-Level Evidence}

The instance-level branch operates on deterministically sampled table rows.
For each table pair, it builds a shared TF--IDF corpus from the sampled values of all source and target columns and compares the value documents associated with each column pair.
The resulting score $h^{\mathrm{inst}}(s,t)\in[0,1]$ captures value-level overlap while down-weighting values that occur broadly across columns.
If a table contains more rows than the configured sample budget, sampling uses a fixed seed so that the evidence is reproducible.

\subsubsection{Evidence Aggregation and Selection}

The enabled name and instance signals are combined by a fixed weighted average,
\[
\widetilde{h}(s,t)
=
\frac{
w_{\mathrm{name}}h^{\mathrm{name}}(s,t)
+w_{\mathrm{inst}}h^{\mathrm{inst}}(s,t)
}{w_{\mathrm{name}}+w_{\mathrm{inst}}}.
\]
The canonical configuration assigns equal internal weight to the two branches.

The operator computes $\widetilde{h}$ over the table pair and then applies bidirectional multiple selection.
A pair is retained only if its score satisfies the configured Top-$n$, relative-to-best, and absolute-threshold conditions from both the source and target directions.
The selected score defines
\[
h(s,t)=\widetilde{h}(s,t),
\]
whereas a localized pair not selected by the complementary operator receives $h(s,t)=0$.
Here, zero indicates that the selection procedure retained no positive complementary evidence for the pair.
The complementary operator supplies scores only and never adds candidates to or removes candidates from $C_k(s)$.

\subsection{Evidence Integration}
\label{sec:fusion}

Decision evidence captures semantic correspondence, whereas the complementary score captures lexical and value regularities; neither signal is uniformly sufficient across heterogeneous schema pairs.
\textsc{JevNexus} combines them into the integrated score
\begin{equation}
f(s,t)=\alpha\,d(s,t)+(1-\alpha)\,h(s,t),
\qquad 0\leq\alpha\leq1.
\label{eq:fusion}
\end{equation}

The coefficient $\alpha$ controls the relative contribution of the two signals.
Because $h(s,t)$ is a matching score rather than a probability, the fused value $f(s,t)$ is used for ranking rather than interpreted as a match probability.

Sorting the localized candidates by $f(s,t)$ produces the default fused order
\[
\pi_s^{f}
=
\operatorname{sort}_{t\in C_k(s)} f(s,t).
\]
This order is returned directly unless the routing criterion in Section~\ref{sec:listwise} invokes listwise refinement.

\subsection{Evidence-Aware Selective Refinement}
\label{sec:listwise}

Pairwise and complementary evidence resolve most localized rankings, but a subset retains both conflicting signals and closely balanced leading candidates.
\textsc{JevNexus} reserves cross-candidate reasoning for this residual ambiguity instead of applying a listwise operator to every source column.

\subsubsection{Routing Criterion}

Let
\[
t_d(s)=\operatorname*{argmax}_{t\in C_k(s)}d(s,t)
\quad\text{and}\quad
t_h(s)=\operatorname*{argmax}_{t\in C_k(s)}h(s,t)
\]
denote the candidates preferred by the two evidence sources.
All argmax and sorting operations use retrieval order as a deterministic tie breaker.
If $h(s,t)=0$ for every localized candidate, the complementary operator has expressed no positive preference and $t_h(s)$ is treated as unavailable.

Evidence disagreement is defined as
\[
D(s)
=
\mathbf{1}\!\left[
t_h(s)\text{ is available}
\;\wedge\;
t_d(s)\neq t_h(s)
\right].
\]
Disagreement alone does not trigger refinement because fusion may still establish a decisive order.

Let $f_{(1)}(s)\geq f_{(2)}(s)$ be the two largest fused scores and define their margin
\[
\Delta_f(s)=f_{(1)}(s)-f_{(2)}(s).
\]
A small margin indicates that the leading fused alternatives remain difficult to distinguish.
The routing decision is therefore
\begin{equation}
g(s)
=
\mathbf{1}\!\left[
D(s)=1
\;\wedge\;
\Delta_f(s)<\tau
\right],
\label{eq:gate}
\end{equation}
where $\tau$ is a fixed uncertainty threshold.
For an empty or singleton candidate set, $\Delta_f(s)$ is undefined and $g(s)$ is set to zero.

\subsubsection{Local Listwise Refinement}

When $g(s)=1$, \textsc{JevNexus} selects the $m_s$ highest-scoring fused candidates,
\[
A_{m_s}(s)
=
\operatorname{TopM}_{t\in C_k(s)}f(s,t),
\qquad
m_s=\min\bigl(m,|C_k(s)|\bigr).
\]
The local refiner receives the source profile as its query and the profiles of $A_{m_s}(s)$ as candidate documents.
Each candidate document also includes its decision, complementary, and fused evidence values so that the refiner can compare the alternatives in the context of the signals that produced the default order.

The refiner returns a permutation $\rho_s$ of $A_{m_s}(s)$ rather than a replacement score scale.
To preserve the scale and distribution of the integrated evidence, let
\[
u_1\geq u_2\geq\cdots\geq u_{m_s}
\]
be the fused score slots occupied by the selected candidates.
If $\rho_s$ places candidate $t$ at refined position $r$, the final localized score of $t$ is set to $u_r$.
Candidates outside $A_{m_s}(s)$ retain their fused scores.

When $g(s)=0$, the fused order bypasses the refiner unchanged.
When the gate opens but the returned permutation agrees with the fused order, the ranking also remains unchanged.
Thus, the refinement stage can modify only the local head of a routed source and cannot change candidate membership or the overall fused score distribution.

\subsection{Complete Ranking}
\label{sec:complete_ranking}

After optional refinement, the candidates in $C_k(s)$ are sorted by their resulting scores to form the localized head.
Target columns outside $C_k(s)$ are appended in their retained target-schema order, yielding the complete ranking $\pi_s$ over $T$.
This construction avoids inventing decision or complementary scores for pairs that the corresponding operators did not evaluate.

Score ties in the localized head are resolved by retrieval order, while the untouched tail preserves target-schema order.
An empty candidate set therefore returns the untouched target order, and a singleton set bypasses listwise refinement by construction.
A ground-truth target outside $C_k(s)$ remains in the tail and cannot be recovered by downstream decision, integration, or refinement.

\subsection{End-to-End Procedure}

Given $S$, $T$, the candidate bound $k$, fusion coefficient $\alpha$, refinement width $m$, and margin threshold $\tau$, \textsc{JevNexus} executes the following procedure for every source column $s\in S$:
\begin{enumerate}
    \item Localize at most $k$ candidates $C_k(s)$ and retain the remaining targets as an ordered tail.
    \item Submit one typed correspondence question for every $t\in C_k(s)$ and collect the decision evidence $d(s,t)$.
    \item Look up the complementary schema/instance evidence $h(s,t)$ and compute the integrated score $f(s,t)$ using Equation~\ref{eq:fusion}.
    \item Compute the evidence-disagreement indicator and fused Top-2 margin, and evaluate the routing rule in Equation~\ref{eq:gate}.
    \item If the gate opens, obtain a local permutation of the fused Top-$m$ and reassign their fused score slots; otherwise, retain the fused order.
    \item Sort the localized candidates, append the untouched tail, and return the complete ranking $\pi_s$.
\end{enumerate}

The complementary score matrix is computed once per table pair, whereas localization, decision questions, routing, and optional refinement are organized per source column.
All neural and non-neural operators remain fixed during online matching, and the procedure performs no online parameter updates.

\subsubsection{Running Example}

Consider again the schematic \texttt{figo\_grade} example from Figure~\ref{fig:motivating-example}.
Localization first places \texttt{tumor\_grade} and \texttt{figo\_stage} in $C_k(s)$ together with any other retrieved candidates.
The typed decision operator asks a separate correspondence question for each candidate, as shown in Figure~\ref{fig:state-question}.
In this schematic example, suppose that attribute semantics yield
\[
d(s,\texttt{tumor\_grade})
>
d(s,\texttt{figo\_stage}),
\]
while the shared token \texttt{figo} makes the complementary evidence prefer the opposite candidate,
\[
h(s,\texttt{tumor\_grade})
<
h(s,\texttt{figo\_stage}).
\]
After integration, suppose that the two candidates occupy the leading fused positions and satisfy $\Delta_f(s)<\tau$.
Because their evidence-specific Top-1 candidates disagree, $D(s)=1$ and the conjunction in Equation~\ref{eq:gate} opens the refinement gate.
In the illustrated path, the local refiner compares only the fused Top-$m$ alternatives and places \texttt{tumor\_grade} ahead of \texttt{figo\_stage} by permuting their fused score slots.
The remaining localized candidates keep their fused scores, and all targets outside $C_k(s)$ are appended in their retained target-schema order.
If either evidence agreement or a decisive fused margin had been present instead, the same example would have bypassed refinement and returned the fused order directly.

\subsection{Inference Cost}

Localization bounds expensive semantic judgment to at most $k$ candidates per source rather than applying it to all $|T|$ target columns.
Across a schema pair, the number of typed questions is at most
\[
\sum_{s\in S}|C_k(s)|\leq |S|k.
\]
Because the questions for a source are batched into one decision request, the normal request count is proportional to $|S|$ rather than to $|S|k$.
The decision interface returns bounded values directly and does not incur autoregressive generation of candidate identifiers, JSON syntax, or numeric score strings.

The complementary operator computes its table-pair score matrix once and exposes the selected entries needed for localized fusion.
Its scores are obtained without generative inference, and lookup during fusion is constant time per localized pair.

Listwise refinement adds exactly
\[
\sum_{s\in S}g(s)
\]
model requests, each over at most $m$ candidates.
An unconditional design would issue one such request for every eligible source column, whereas \textsc{JevNexus} pays this additional cost only when evidence disagreement coincides with a small fused margin.
The canonical values of $k$, $\alpha$, $m$, and $\tau$, together with the concrete operator implementations, are reported in Section~\ref{sec:setup}.

\section{Experimental Evaluation}
\label{sec:evaluation}

We evaluate \textsc{JevNexus} along four dimensions: end-to-end effectiveness,
inference efficiency, component contributions, and the behavior of selective
refinement. We additionally study hyperparameter sensitivity and decompose
residual errors into candidate-localization and downstream-ranking failures.

\subsection{Experimental Setup}
\label{sec:setup}

\paragraph{Datasets.}
We use GDC-SM~\cite{liu2025magneto} and five datasets distributed with
Valentine~\cite{koutras2021valentine}. Table~\ref{tab:dataset-statistics}
summarizes the benchmark. It contains 561 cases, 13,522 source columns, and
8,944 ground-truth pairs. GDC is particularly challenging because each case
maps into a common target schema with 736 columns.

\begin{table}[t]
\centering
\caption{Dataset statistics.}
\label{tab:dataset-statistics}
\setlength{\tabcolsep}{3.2pt}
\resizebox{\columnwidth}{!}{%
\begin{tabular}{lrrrrrr}
\toprule
Dataset & Cases & Source cols. & Source width & Target width & GT pairs & GT/case \\
\midrule
ChEMBL & 180 & 3,060 & 12--23 & 12--23 & 2,052 & 11.4 \\
GDC & 10 & 569 & 16--179 & 736 & 259 & 25.9 \\
Magellan & 7 & 41 & 4--9 & 4--9 & 41 & 5.9 \\
OpenData & 180 & 6,876 & 26--51 & 26--51 & 4,572 & 25.4 \\
TPC-DI & 180 & 2,916 & 11--22 & 12--22 & 1,980 & 11.0 \\
WikiData & 4 & 60 & 13--20 & 13--20 & 40 & 10.0 \\
\midrule
Total & 561 & 13,522 & -- & -- & 8,944 & 15.9 \\
\bottomrule
\end{tabular}
}
\end{table}

\paragraph{Compared methods.}
We compare against COMA~\cite{do2002coma}, COMA++~\cite{engmann2007comapp}, a
distribution-based matcher, Similarity Flooding~\cite{melnik2002similarity},
ISResMat~\cite{du2024isresmat}, Unicorn~\cite{tu2023unicorn}, and
\textsc{Magneto}~\cite{liu2025magneto}. We run the official \textsc{Magneto}
implementation and instantiate its replaceable LLM reranker with Qwen3.5-9B
\cite{qwen2026qwen35}. \textsc{Magneto} and \textsc{JevNexus} receive the same
localized candidate sets and use the same candidate limit, isolating the
quality and cost of downstream matching from candidate recall. The ablations retain
only typed decision evidence, add complementary schema/instance evidence,
apply listwise refinement unconditionally, or use the complete selective
pipeline.

\paragraph{Metrics and protocol.}
MRR is our primary ranking metric, Hits@1 measures whether a valid target is
ranked first, and Recall@GT follows the Valentine and \textsc{Magneto}
protocol: all pairs in a case are pooled by score, the top $|G|$ are retained,
and recall is computed against the ground truth. Recall@GT therefore also
depends on score comparability across source columns.

We report two aggregate views. \emph{Case-macro} averages all cases equally;
therefore ChEMBL, OpenData, and TPC-DI, with 180 cases each, receive more
weight. \emph{Dataset-macro} first averages cases within each dataset and then
gives each dataset equal weight. \textsc{JevNexus} and all ablations complete all
561 cases. \textsc{Magneto} times out on one OpenData case, and Distribution
does not complete the ten GDC cases. Main effectiveness results cover all 561
expected cases and assign zero effectiveness to each execution failure;
completed-case counts are reported explicitly.

\paragraph{Configuration.}
The candidate limit is $k=20$. The complementary matcher uses $n=20$, a
relative-to-best tolerance of $0.15$, an absolute threshold of $0$, equal
schema and instance weights, and an instance sample size of 500. Decision
inference uses Open-Jev-9B with maximum length 4096 and batch size
20~\cite{cai2026openjev}. Fusion uses
$\alpha=0.400$, and refinement uses $m=3$, $\tau=0.020$, and
Jina-Reranker-v3.5~\cite{nasika2026jina}. The same fusion and routing
configuration is used for every dataset without per-dataset fitting.
\textsc{Magneto} uses Qwen3.5-9B with temperature zero, at most 1,024 generated
tokens, and disabled thinking. We retain its original benchmark serialization
defaults but use the zero-shot retriever rather than the GDC-specific
fine-tuned variant. Neural methods run
on the same RTX PRO 6000 host; timing excludes model startup and includes
representation, retrieval, matching or reranking, and final ranking. All
system-level latency summaries include the elapsed time of a failed run.
All
metrics are computed from saved rankings with the same evaluator. Exact model
revisions, prompts, software versions, and commands are provided in the
artifact. Code and experimental artifacts are available at
\url{https://github.com/RazeenLI/JevNexus}.

\subsection{End-to-End Effectiveness}
\label{sec:effectiveness}

\begin{figure}[t]
\centering
\includegraphics[width=\columnwidth]{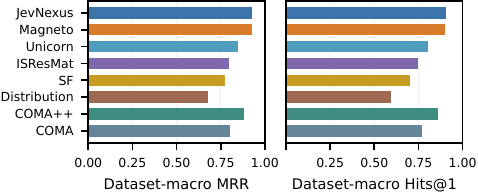}
\caption{Overall matching effectiveness.}
\label{fig:headline-results}
\end{figure}

Figure~\ref{fig:headline-results} summarizes performance across all six
benchmark families. \textsc{JevNexus} obtains the highest dataset-macro MRR and
Hits@1, followed by \textsc{Magneto}, while \textsc{Magneto} obtains the
highest Recall@GT. The remaining methods trail both on aggregate ranking
quality, although their relative positions vary by dataset.

Table~\ref{tab:main-results} provides the per-dataset results; each dataset
block reports MRR, Hits@1, and Recall@GT, with the best values in bold. \textsc{JevNexus}
achieves the highest MRR and Hits@1 on GDC, ties for the best ranking results
on Magellan and WikiData, and is close to the leading method on OpenData and
TPC-DI. ChEMBL is the only family on which its MRR trails \textsc{Magneto} by
more than two points, although it remains stronger than every other baseline.
No method dominates every setting: Unicorn leads MRR and Hits@1 on OpenData,
while \textsc{Magneto} leads on ChEMBL and narrowly on TPC-DI. Recall@GT gives
a different view, with \textsc{Magneto} strongest on four datasets and
\textsc{JevNexus} strongest on WikiData.

\begin{table*}[t]
\centering
\caption{Per-dataset matching effectiveness.}
\label{tab:main-results}
\resizebox{\textwidth}{!}{%
\begin{tabular}{l*{6}{rrr}}
\toprule
& \multicolumn{3}{c}{ChEMBL} & \multicolumn{3}{c}{GDC} & \multicolumn{3}{c}{Magellan} & \multicolumn{3}{c}{OpenData} & \multicolumn{3}{c}{TPC-DI} & \multicolumn{3}{c}{WikiData} \\
\cmidrule(lr){2-4}\cmidrule(lr){5-7}\cmidrule(lr){8-10}\cmidrule(lr){11-13}\cmidrule(lr){14-16}\cmidrule(lr){17-19}
Method & MRR & H@1 & R@GT & MRR & H@1 & R@GT & MRR & H@1 & R@GT & MRR & H@1 & R@GT & MRR & H@1 & R@GT & MRR & H@1 & R@GT \\
\midrule
COMA & 0.790 & 0.715 & 0.599 & 0.521 & 0.481 & 0.314 & \textbf{1.000} & \textbf{1.000} & \textbf{1.000} & 0.814 & 0.781 & 0.641 & 0.903 & 0.881 & 0.876 & 0.784 & 0.748 & 0.665 \\
COMA++ & 0.896 & 0.847 & 0.747 & 0.561 & 0.507 & 0.366 & \textbf{1.000} & \textbf{1.000} & \textbf{1.000} & 0.911 & 0.891 & 0.514 & 0.972 & 0.963 & 0.870 & \textbf{0.946} & \textbf{0.946} & 0.873 \\
Distribution & 0.549 & 0.438 & 0.451 & 0.000 & 0.000 & 0.000 & 0.686 & 0.561 & 0.561 & 0.563 & 0.482 & 0.282 & 0.790 & 0.722 & 0.584 & 0.808 & 0.773 & 0.644 \\
Similarity Flooding & 0.793 & 0.659 & 0.447 & 0.544 & 0.467 & 0.280 & \textbf{1.000} & \textbf{1.000} & \textbf{1.000} & 0.650 & 0.554 & 0.420 & 0.877 & 0.825 & 0.665 & 0.766 & 0.710 & 0.646 \\
ISResMat & 0.838 & 0.782 & 0.733 & 0.448 & 0.379 & 0.258 & \textbf{1.000} & \textbf{1.000} & \textbf{1.000} & 0.686 & 0.587 & 0.514 & 0.868 & 0.794 & 0.767 & \textbf{0.946} & \textbf{0.946} & 0.873 \\
Unicorn & 0.814 & 0.736 & 0.684 & 0.578 & 0.494 & 0.319 & 0.963 & 0.944 & 0.920 & \textbf{0.971} & \textbf{0.955} & \textbf{0.793} & 0.872 & 0.817 & 0.651 & 0.900 & 0.877 & 0.817 \\
\textsc{Magneto} & \textbf{0.965} & \textbf{0.944} & \textbf{0.883} & 0.738 & 0.657 & \textbf{0.461} & \textbf{1.000} & \textbf{1.000} & \textbf{1.000} & 0.932 & 0.912 & 0.753 & \textbf{0.983} & \textbf{0.968} & \textbf{0.945} & 0.940 & 0.933 & 0.890 \\
\textbf{JevNexus} & 0.907 & 0.878 & 0.783 & \textbf{0.794} & \textbf{0.725} & 0.456 & \textbf{1.000} & \textbf{1.000} & \textbf{1.000} & 0.952 & 0.939 & 0.648 & 0.980 & 0.964 & 0.865 & \textbf{0.946} & \textbf{0.946} & \textbf{0.902} \\
\bottomrule
\end{tabular}
}
\end{table*}

\paragraph{Comparison with generative reranking.}
Because \textsc{Magneto} is the strongest generative method and the closest
overall competitor, we additionally bootstrap paired case differences within
each dataset before averaging the six dataset means. The analysis includes all
561 cases and assigns zero effectiveness to the timed-out \textsc{Magneto}
case. The MRR difference (\textsc{JevNexus}$-$\textsc{Magneto}) is 0.004 with a
95\% confidence interval of $[-0.005,0.012]$; the Hits@1 difference is 0.006
with interval $[-0.007,0.020]$. Both intervals contain zero. The Recall@GT
difference is $-0.046$ with interval $[-0.061,-0.032]$ ($p<0.001$). Thus,
their dataset-macro MRR and Hits@1 are statistically comparable, whereas
\textsc{Magneto} retains higher pooled correspondence recall.

\subsection{Efficiency}
\label{sec:efficiency}

\begin{figure}[t]
\centering
\includegraphics[width=\columnwidth]{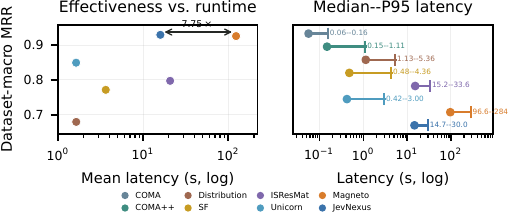}
\caption{Matching effectiveness and inference efficiency.}
\label{fig:efficiency}
\end{figure}

The left panel of Figure~\ref{fig:efficiency} places \textsc{JevNexus} in the
effectiveness--runtime landscape of methods evaluated across all six benchmark
families, while the right panel compares median and tail latency for all
methods. Traditional matchers
are faster but substantially less accurate, while \textsc{JevNexus} combines the
highest dataset-macro MRR with much lower latency than \textsc{Magneto} and
ISResMat. Including the timed-out run, \textsc{JevNexus} requires 15.929 seconds
per case compared with 123.452 seconds for \textsc{Magneto}, a
7.750$\times$ speedup.

Median and 95th-percentile latency are 14.694 and 30.029 seconds for
\textsc{JevNexus}, versus 96.639 and 284.221 seconds for \textsc{Magneto}.
The decision pipeline processes more input context but emits no autoregressive
score tokens, whereas the completed \textsc{Magneto} runs emit 5.31 million
recorded output tokens. The bounded output interface removes autoregressive
score generation, and the latency advantage persists across both typical and
tail cases.

\subsection{Component Ablation}
\label{sec:component_ablation}

Table~\ref{tab:component-ablation} uses the same three metrics as the main
table. Adding complementary schema/instance evidence to typed decisions
produces the largest consistent gain. Unconditional refinement is harmful on
ChEMBL, TPC-DI, and WikiData. Selective refinement avoids this instability and
produces the strongest aggregate variant, although its improvement over the
unrefined fusion is deliberately modest.

\begin{table*}[t]
\centering
\caption{Component ablation by dataset.}
\label{tab:component-ablation}
\resizebox{\textwidth}{!}{%
\begin{tabular}{l*{6}{rrr}}
\toprule
& \multicolumn{3}{c}{ChEMBL} & \multicolumn{3}{c}{GDC} & \multicolumn{3}{c}{Magellan} & \multicolumn{3}{c}{OpenData} & \multicolumn{3}{c}{TPC-DI} & \multicolumn{3}{c}{WikiData} \\
\cmidrule(lr){2-4}\cmidrule(lr){5-7}\cmidrule(lr){8-10}\cmidrule(lr){11-13}\cmidrule(lr){14-16}\cmidrule(lr){17-19}
Variant & MRR & H@1 & R@GT & MRR & H@1 & R@GT & MRR & H@1 & R@GT & MRR & H@1 & R@GT & MRR & H@1 & R@GT & MRR & H@1 & R@GT \\
\midrule
w/o Both & 0.830 & 0.766 & 0.616 & 0.739 & 0.642 & 0.376 & 0.944 & 0.895 & 0.855 & 0.918 & 0.882 & 0.620 & 0.961 & 0.931 & 0.774 & 0.868 & 0.817 & 0.773 \\
w/o Complement & 0.869 & 0.832 & 0.638 & 0.797 & \textbf{0.751} & 0.400 & 0.990 & 0.980 & 0.911 & 0.948 & 0.935 & 0.626 & 0.941 & 0.903 & 0.762 & 0.889 & 0.858 & 0.773 \\
w/o Refinement & 0.904 & 0.873 & 0.783 & 0.781 & 0.703 & 0.456 & \textbf{1.000} & \textbf{1.000} & \textbf{1.000} & 0.950 & 0.935 & \textbf{0.647} & 0.980 & 0.963 & \textbf{0.865} & \textbf{0.946} & \textbf{0.946} & \textbf{0.902} \\
Always Refine & 0.901 & 0.865 & 0.762 & \textbf{0.798} & 0.730 & 0.454 & \textbf{1.000} & \textbf{1.000} & \textbf{1.000} & 0.952 & 0.939 & 0.645 & 0.935 & 0.889 & 0.829 & 0.924 & 0.902 & \textbf{0.902} \\
\textbf{JevNexus} & \textbf{0.907} & \textbf{0.878} & \textbf{0.783} & 0.794 & 0.725 & \textbf{0.456} & \textbf{1.000} & \textbf{1.000} & \textbf{1.000} & \textbf{0.952} & \textbf{0.939} & \textbf{0.648} & \textbf{0.980} & \textbf{0.964} & \textbf{0.865} & \textbf{0.946} & \textbf{0.946} & \textbf{0.902} \\
\bottomrule
\end{tabular}
}
\end{table*}

Case-macro MRR for removing both components, removing complementary evidence,
removing selective refinement, always refining, and \textsc{JevNexus} is
0.900, 0.918, 0.942, 0.928, and 0.944, respectively. Corresponding Hits@1 is
0.856, 0.888, 0.921, 0.896, and 0.925; Recall@GT is 0.668, 0.674, 0.763,
0.744, and 0.764. Selective refinement is therefore a targeted repair
mechanism, not the primary source of accuracy.

\subsection{Selective Routing}
\label{sec:routing}

Table~\ref{tab:routing} reports the online gate rather than an offline replay.
It opens for 766 of 13,522 source columns (5.665\%), avoiding 94.335\% of the
listwise calls made by unconditional refinement. OpenData and GDC contain the highest proportions
of conflicted, low-margin decisions; the gate never opens on Magellan or
WikiData.

\begin{table}[t]
\centering
\caption{Observed online routing by dataset.}
\label{tab:routing}
\setlength{\tabcolsep}{3.1pt}
\resizebox{\columnwidth}{!}{%
\begin{tabular}{lrrrrrr}
\toprule
Dataset & Sources & Disagree & Low margin & Calls & Avoided & Activation (\%) \\
\midrule
ChEMBL & 3,060 & 711 & 717 & 111 & 2,949 & 3.627 \\
GDC & 569 & 179 & 189 & 38 & 531 & 6.678 \\
Magellan & 41 & 4 & 0 & 0 & 41 & 0.000 \\
OpenData & 6,876 & 1,596 & 1,515 & 521 & 6,355 & 7.577 \\
TPC-DI & 2,916 & 338 & 786 & 96 & 2,820 & 3.292 \\
WikiData & 60 & 9 & 7 & 0 & 60 & 0.000 \\
\midrule
Overall & 13,522 & 2,837 & 3,214 & 766 & 12,756 & 5.665 \\
\bottomrule
\end{tabular}
}
\end{table}

\begin{figure}[t]
\centering
\includegraphics[width=\columnwidth]{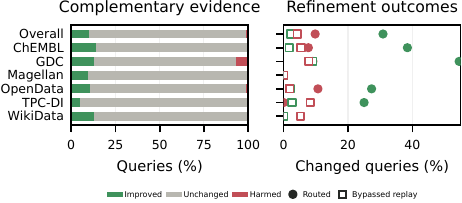}
\caption{Effects of evidence integration and selective refinement.}
\label{fig:mechanism-analysis}
\end{figure}

Figure~\ref{fig:mechanism-analysis} tests whether the two mechanisms change the
right queries. Complementary evidence improves the reciprocal rank of 10.262\%
of ground-truth-bearing queries and harms 0.995\%. Among the 204 routed source
columns with at least one ground-truth correspondence, refinement improves the
ranking for 30.882\% and degrades it for 9.804\%.
Conversely, replaying unconditional refinement on bypassed queries harms
4.257\% but improves only 2.175\%. The gate therefore concentrates refinement
where it is more likely to help while protecting comparatively stable rankings.

The avoided calls are not the primary source of the speedup over
\textsc{Magneto}; the overall comparison contrasts typed decision inference
with autoregressive generation. Within \textsc{JevNexus}, unconditional refinement takes
16.242 seconds per case and selective refinement takes 15.929 seconds. The
gate's main empirical role is therefore ranking control, with a smaller
secondary runtime benefit.

\subsection{Sensitivity Analysis}
\label{sec:sensitivity}

\begin{figure*}[t]
\centering
\includegraphics[width=\textwidth]{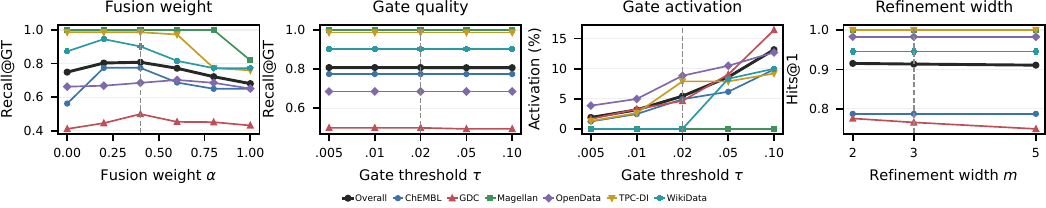}
\caption{Hyperparameter sensitivity on the development subset.}
\label{fig:sensitivity}
\end{figure*}

Figure~\ref{fig:sensitivity} reports one-factor-at-a-time diagnostic sweeps on
the fixed 29-case development subset. Overall denotes the dataset-macro
average; the sweeps characterize, rather than select, the reported setting.

\emph{Fusion and gate.} Overall Recall@GT rises from 0.749 at $\alpha=0.000$
to 0.808 at $\alpha=0.400$, then declines. The per-dataset curves show that no
single weight dominates: GDC peaks at 0.400, whereas OpenData peaks at 0.600.
Increasing $\tau$ from 0.020 to 0.100 raises activation from 5.443\% to
13.165\%, but does not improve Overall Recall@GT (0.808 versus 0.807).

\emph{Refinement width.} For $m=2,3,5$, Overall Hits@1 is 0.915, 0.914, and
0.911. Only GDC changes visibly (0.776, 0.766, and 0.749); the other datasets
are unchanged. Thus the canonical $\alpha=0.400$, $\tau=0.020$, and $m=3$ is a
fixed low-routing operating point, not a sharply tuned optimum.

\subsection{Error Analysis}
\label{sec:error_analysis}

\begin{figure}[t]
\centering
\includegraphics[width=\columnwidth]{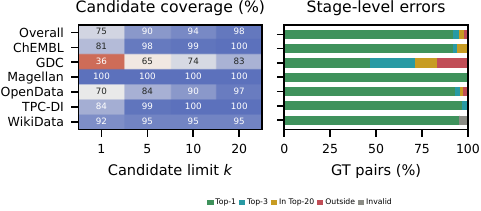}
\caption{Candidate coverage and stage-wise error decomposition.}
\label{fig:diagnostics}
\end{figure}

Figure~\ref{fig:diagnostics} separates candidate coverage from downstream
ordering. Candidate coverage rises from 74.832\% at $k=1$ to 98.032\% at
$k=20$. At the reported operating point, 92.039\% of
ground-truth pairs are ranked first, 3.321\% reach the final Top-3 but not rank
one, 2.661\% enter the candidate Top-20 but remain below the final Top-3, and
1.957\% fall outside the candidate set. The remaining 0.022\% are the two
WikiData pairs whose target column is absent from the raw target table.

GDC is the most difficult benchmark for nearly all evaluated methods because
each source table is matched against a common 736-column target schema.
Nevertheless, \textsc{JevNexus} achieves the highest GDC MRR and Hits@1 among the
compared methods and nearly matches the best Recall@GT. We therefore examine
the errors that remain at this comparatively strong operating point. On GDC,
46.718\% of its ground-truth pairs are ranked first,
24.324\% are in the final Top-3 but below rank one, 12.355\% are in the
candidate Top-20 but below the final Top-3, and 16.602\% never enter the
candidate set. Its 736-column target schema exposes both a localization ceiling
and a local-ordering problem. By contrast, Magellan is saturated, and TPC-DI
has no Top-20 localization misses.

The decomposition does not by itself explain the Recall@GT difference. Strong
MRR and Hits@1 can coexist with weaker Recall@GT because the latter pools
scores across source columns. The observed pattern is consistent with a
cross-query score-comparability issue, but establishing calibration as its
cause requires a dedicated calibration experiment; we therefore leave that
claim to further analysis rather than infer it from ranking metrics alone.

\section{Related Work}
\label{sec:related}

\paragraph{Traditional and Composite Schema Matching.}
Classical schema matching derives correspondences from names, constraints,
schema structure, and data instances. Rahm and Bernstein organize this design
space along schema- versus instance-level, element- versus structure-level,
and linguistic versus constraint-based evidence~\cite{rahm2001survey}. Cupid
combines linguistic and structural similarity~\cite{madhavan2001cupid}, COMA
provides an extensible architecture for composing multiple
matchers~\cite{do2002coma}, and Similarity Flooding propagates evidence through
graph structure~\cite{melnik2002similarity}. COMA++ and later evaluation in
Valentine further demonstrate the value of combining heterogeneous signals
and the sensitivity of classical matchers to data and workload
characteristics~\cite{engmann2007comapp,koutras2021valentine}. JevNexus retains
this central insight---matching evidence is complementary---but combines it
with neural semantic judgments and separates evidence fusion from optional
cross-candidate reasoning.

\paragraph{Learned Schema Matching.}
Learning-based methods replace or adjust hand-designed similarities using
representations learned from data. EmbDI constructs local relational
embeddings for schema matching and entity resolution~\cite{cappuzzo2020embdi},
whereas ADnEV learns to adjust and evaluate similarity matrices produced by
existing matchers across domains~\cite{shraga2020adnev}. More recent systems
use pretrained language models: LSM combines a fine-tuned encoder with active
learning for data-free matching against large target
schemas~\cite{zhang2023plm}; Unicorn trains a unified model across multiple
data-matching tasks~\cite{tu2023unicorn}; and ISResMat adapts a pretrained
encoder in situ using schema and instance evidence~\cite{du2024isresmat}.
These systems principally improve representations, training, or transfer.
JevNexus instead studies the inference architecture after candidate localization:
how semantic correspondence should be represented, combined with
complementary evidence, and selectively compared.

\paragraph{LLM-Based Schema Matching.}
Foundation models have been prompted for a range of data cleaning and
integration tasks, establishing their potential for semantic matching without
task-specific architectures~\cite{narayan2022foundation,zhang2024jellyfish,seedat2024matchmaker,sheetrit2024rematch,chen2026construm}. For schema matching,
Parciak et al. study how prompt scope and available context affect generated
correspondences, decisiveness, and complementarity with lexical
similarity~\cite{parciak2024schema}. Magneto introduces the closest
retrieve--rerank architecture: an efficient semantic model localizes candidate
columns before a generative LLM assigns scores and reranks the localized
set~\cite{liu2025magneto}. JevNexus adopts candidate localization as a computational
abstraction, but changes the main inference interface from generated scores to
bounded pairwise decisions, explicitly fuses complementary evidence, and
reserves joint listwise comparison for ambiguous candidate sets.

\paragraph{Decision-Oriented and Selective Reranking.}
Related matching systems commonly cast correspondence as a bounded
classification problem; for example, Ditto formulates entity matching as
sequence-pair classification with pretrained language
models~\cite{li2020ditto}, while Unicorn applies a shared matching interface
across heterogeneous data-matching tasks~\cite{tu2023unicorn}. Ranking research
further distinguishes independent scoring from joint listwise comparison~\cite{cao2007listnet,xia2008listmle}.
RankT5 directly produces ranking scores under pairwise or listwise
losses~\cite{zhuang2023rankt5}, whereas RankGPT and ListT5 use language models
to order candidates jointly~\cite{sun2023rankgpt,yoon2024listt5}. FIRST reduces
the autoregressive cost of listwise LLM reranking by deriving an ordering from
the logits of a single generated identifier~\cite{reddy2024first}. More
broadly, RouteLLM learns to invoke a stronger model selectively to balance
quality and inference cost~\cite{schuster2022calm,gupta2024cascades,dekoninck2025routing,ong2025routellm}. JevNexus brings these concerns
together for schema matching: it uses direct bounded decisions on the main
path and an evidence-based, training-free gate to invoke local listwise
refinement only when candidate-level comparison is warranted.

\section{Discussion and Limitations}
\label{sec:discussion}

\paragraph{Effectiveness--efficiency trade-off.}
Decision-centric matching retains competitive ranking effectiveness while
substantially reducing the cost of semantic inference. Under a shared
candidate generator, \textsc{JevNexus} is statistically comparable to the strongest
tested generative method in dataset-macro MRR and Hits@1, while reducing mean
online runtime by $7.750\times$. The ablations indicate that complementary
schema and instance evidence contributes the main consistent gain, confirming
that semantic representations do not subsume all useful matching signals.
Listwise reasoning has a narrower role: applying it uniformly can degrade the
ranking, whereas selective refinement is useful when evidence conflicts and
the leading candidates remain close. The resulting separation makes pairwise
decisions the efficient default and reserves joint comparison for unresolved
cross-candidate ambiguities.

\paragraph{Localization and score comparability.}
The bounded candidate set remains a hard limit on downstream matching. If a
ground-truth target is absent from $C_k(s)$, neither pairwise decisions nor
listwise refinement can recover it. Simply increasing $k$ may improve coverage,
but also increases inference cost and exposes the matcher to more distractors.
An important direction is therefore adaptive localization that expands the
candidate set only when the available evidence indicates that the initial set
is insufficient. A separate limitation concerns the use of decision and fused
scores across source queries. The strong query-level ranking results show that
these scores are effective for ordering candidates for the same source, but
their weaker pooled correspondence recall suggests limited comparability
across source queries, motivating direct evaluation and calibration of
cross-query scores.

\paragraph{Transferability of fusion and routing.}
The current system uses fixed values for the fusion coefficient, refinement
width, and routing margin. Sensitivity analysis identifies a stable,
low-routing operating point, but also shows that no single setting dominates
across all benchmark families. This variation is expected because datasets
differ in the informativeness of column names and values, the density of
plausible candidates, and the width of their target schemas. The shared
configuration provides a stable operating point across datasets, while
learning evidence weights, adapting candidate and refinement widths per query,
and routing from the full pattern of evidence remain promising directions.

\paragraph{Evaluation and system scope.}
The evaluation spans six heterogeneous benchmark families, but it does not
cover every schema-matching regime. In particular, several families contain
relatively few cases, and the experiments use one decision backbone and one
listwise refinement model. Broader evaluation across unseen domains, model
families, and substantially larger target schemas is therefore needed to
establish how far the observed trade-off transfers. Absolute latency depends
on hardware, serving infrastructure, and batching; our measurements therefore
characterize relative efficiency under a shared experimental environment.

\section{Conclusion}
\label{sec:conclusion}

We presented \textsc{JevNexus}, a decision-centric approach to schema matching that combines typed pairwise judgments with complementary schema and instance evidence, and invokes listwise reasoning only for uncertain cases. This design preserves localized correspondence evidence while avoiding the cost and occasional degradation associated with unconditional generative reranking.

Across six benchmark families, \textsc{JevNexus} achieves MRR and Hits@1 statistically comparable to those of \textsc{Magneto} while reducing mean online runtime by $7.750\times$, although \textsc{Magneto} attains higher Recall@GT. The experiments further show that evidence fusion provides a consistent benefit and that selective refinement is most useful when the evidence sources disagree and the resulting ranking remains uncertain. Retrieval coverage remains the principal limitation: refinement cannot recover a correct correspondence absent from the candidate set. Future work will therefore investigate adaptive candidate expansion and calibration across source queries, together with broader evaluation across domains and execution environments.

\bibliographystyle{ACM-Reference-Format}
\bibliography{references}

@article{rahm2001survey,
  author  = {Erhard Rahm and Philip A. Bernstein},
  title   = {A Survey of Approaches to Automatic Schema Matching},
  journal = {The VLDB Journal},
  volume  = {10},
  number  = {4},
  pages   = {334--350},
  year    = {2001},
  doi     = {10.1007/s007780100057}
}

@article{li2023proteogenomic,
  author  = {Yize Li and Yongchao Dou and Felipe da Veiga Leprevost and Yifat Geffen and Anna P. Calinawan and Fran{\c{c}}ois Aguet and others},
  title   = {Proteogenomic Data and Resources for Pan-Cancer Analysis},
  journal = {Cancer Cell},
  volume  = {41},
  number  = {8},
  pages   = {1397--1406},
  year    = {2023},
  doi     = {10.1016/j.ccell.2023.06.009}
}

@inproceedings{fernandez2018aurum,
  author    = {Raul Castro Fernandez and Ziawasch Abedjan and Famien Koko and Gina Yuan and Samuel Madden and Michael Stonebraker},
  title     = {Aurum: A Data Discovery System},
  booktitle = {Proceedings of the 34th IEEE International Conference on Data Engineering (ICDE)},
  pages     = {1001--1012},
  year      = {2018},
  doi       = {10.1109/ICDE.2018.00094}
}

@article{cafarella2008webtables,
  author  = {Michael J. Cafarella and Alon Halevy and Daisy Zhe Wang and Eugene Wu and Yang Zhang},
  title   = {WebTables: Exploring the Power of Tables on the Web},
  journal = {Proceedings of the VLDB Endowment},
  volume  = {1},
  number  = {1},
  pages   = {538--549},
  year    = {2008},
  doi     = {10.14778/1453856.1453916}
}

@article{deng2024lakebench,
  author  = {Yuhao Deng and Chengliang Chai and Lei Cao and Qin Yuan and Siyuan Chen and Yanrui Yu and Zhaoze Sun and Junyi Wang and Jiajun Li and Ziqi Cao and Kaisen Jin and Chi Zhang and Yuqing Jiang and Yuanfang Zhang and Yuping Wang and Ye Yuan and Guoren Wang and Nan Tang},
  title   = {LakeBench: A Benchmark for Discovering Joinable and Unionable Tables in Data Lakes},
  journal = {Proceedings of the VLDB Endowment},
  volume  = {17},
  number  = {8},
  pages   = {1925--1938},
  year    = {2024},
  doi     = {10.14778/3659437.3659448}
}

@inproceedings{madhavan2001cupid,
  author    = {Jayant Madhavan and Philip A. Bernstein and Erhard Rahm},
  title     = {Generic Schema Matching with Cupid},
  booktitle = {Proceedings of the 27th International Conference on Very Large Data Bases (VLDB)},
  pages     = {49--58},
  year      = {2001}
}

@inproceedings{do2002coma,
  author    = {Hong-Hai Do and Erhard Rahm},
  title     = {COMA -- A System for Flexible Combination of Schema Matching Approaches},
  booktitle = {Proceedings of the 28th International Conference on Very Large Databases (VLDB)},
  pages     = {610--621},
  year      = {2002},
  doi       = {10.1016/B978-155860869-6/50060-3}
}

@inproceedings{melnik2002similarity,
  author    = {Sergey Melnik and Hector Garcia-Molina and Erhard Rahm},
  title     = {Similarity Flooding: A Versatile Graph Matching Algorithm and Its Application to Schema Matching},
  booktitle = {Proceedings of the 18th IEEE International Conference on Data Engineering (ICDE)},
  pages     = {117--128},
  year      = {2002},
  doi       = {10.1109/ICDE.2002.994702}
}

@inproceedings{engmann2007comapp,
  author    = {Daniel Engmann and Sabine Massmann},
  title     = {Instance Matching with COMA++},
  booktitle = {BTW Workshops},
  volume    = {7},
  pages     = {28--37},
  year      = {2007}
}

@inproceedings{koutras2021valentine,
  author    = {Christos Koutras and George Siachamis and Andra Ionescu and Kyriakos Psarakis and Jerry Brons and Marios Fragkoulis and Christoph Lofi and Angela Bonifati and Asterios Katsifodimos},
  title     = {Valentine: Evaluating Matching Techniques for Dataset Discovery},
  booktitle = {Proceedings of the 37th IEEE International Conference on Data Engineering (ICDE)},
  pages     = {468--479},
  year      = {2021}
}

@inproceedings{zhang2023plm,
  author    = {Yunjia Zhang and Avrilia Floratou and Joyce Cahoon and Subru Krishnan and Andreas C. M{\"u}ller and Dalitso Banda and Fotis Psallidas and Jignesh M. Patel},
  title     = {Schema Matching using Pre-Trained Language Models},
  booktitle = {Proceedings of the 39th IEEE International Conference on Data Engineering (ICDE)},
  pages     = {1558--1571},
  year      = {2023},
  doi       = {10.1109/ICDE55515.2023.00123}
}

@article{tu2023unicorn,
  author  = {Jianhong Tu and Ju Fan and Nan Tang and Peng Wang and Guoliang Li and Xiaoyong Du and Xiaofeng Jia and Song Gao},
  title   = {Unicorn: A Unified Multi-tasking Model for Supporting Matching Tasks in Data Integration},
  journal = {Proceedings of the ACM on Management of Data},
  volume  = {1},
  number  = {1},
  pages   = {84:1--84:26},
  year    = {2023},
  doi     = {10.1145/3588938}
}

@inproceedings{du2024isresmat,
  author    = {Xingyu Du and Gongsheng Yuan and Sai Wu and Gang Chen and Peng Lu},
  title     = {In Situ Neural Relational Schema Matcher},
  booktitle = {Proceedings of the 40th IEEE International Conference on Data Engineering (ICDE)},
  pages     = {138--150},
  year      = {2024},
  doi       = {10.1109/ICDE60146.2024.00018}
}

@article{liu2025magneto,
  author  = {Yurong Liu and Eduardo H. M. Pena and A{\'e}cio S. R. Santos and Eden Wu and Juliana Freire},
  title   = {Magneto: Combining Small and Large Language Models for Schema Matching},
  journal = {Proceedings of the VLDB Endowment},
  volume  = {18},
  number  = {8},
  pages   = {2681--2694},
  year    = {2025},
  doi     = {10.14778/3742728.3742757}
}

@inproceedings{cappuzzo2020embdi,
  author    = {Riccardo Cappuzzo and Paolo Papotti and Saravanan Thirumuruganathan},
  title     = {Creating Embeddings of Heterogeneous Relational Datasets for Data Integration Tasks},
  booktitle = {Proceedings of the 2020 ACM SIGMOD International Conference on Management of Data},
  pages     = {1335--1349},
  year      = {2020},
  doi       = {10.1145/3318464.3389742}
}

@article{shraga2020adnev,
  author  = {Roee Shraga and Avigdor Gal and Haggai Roitman},
  title   = {{ADnEV}: Cross-Domain Schema Matching using Deep Similarity Matrix Adjustment and Evaluation},
  journal = {Proceedings of the VLDB Endowment},
  volume  = {13},
  number  = {9},
  pages   = {1401--1415},
  year    = {2020},
  doi     = {10.14778/3397230.3397237}
}

@article{narayan2022foundation,
  author  = {Avanika Narayan and Ines Chami and Laurel Orr and Christopher R{\'e}},
  title   = {Can Foundation Models Wrangle Your Data?},
  journal = {Proceedings of the VLDB Endowment},
  volume  = {16},
  number  = {4},
  pages   = {738--746},
  year    = {2022},
  doi     = {10.14778/3574245.3574258}
}

@inproceedings{parciak2024schema,
  author    = {Marcel Parciak and Brecht Vandevoort and Frank Neven and Liesbet M. Peeters and Stijn Vansummeren},
  title     = {Schema Matching with Large Language Models: An Experimental Study},
  booktitle = {Proceedings of the VLDB 2024 Workshop on Tabular Data Analysis (TaDA)},
  year      = {2024},
  url       = {https://www.vldb.org/workshops/2024/proceedings/TaDA/TaDA.8.pdf}
}

@article{li2020ditto,
  author  = {Yuliang Li and Jinfeng Li and Yoshihiko Suhara and AnHai Doan and Wang-Chiew Tan},
  title   = {Deep Entity Matching with Pre-Trained Language Models},
  journal = {Proceedings of the VLDB Endowment},
  volume  = {14},
  number  = {1},
  pages   = {50--60},
  year    = {2020},
  doi     = {10.14778/3421424.3421431}
}

@inproceedings{zhuang2023rankt5,
  author    = {Honglei Zhuang and Zhen Qin and Rolf Jagerman and Kai Hui and Ji Ma and Jing Lu and Jianmo Ni and Xuanhui Wang and Michael Bendersky},
  title     = {{RankT5}: Fine-Tuning {T5} for Text Ranking with Ranking Losses},
  booktitle = {Proceedings of the 46th International ACM SIGIR Conference on Research and Development in Information Retrieval},
  pages     = {2308--2313},
  year      = {2023},
  doi       = {10.1145/3539618.3592047}
}

@inproceedings{sun2023rankgpt,
  author    = {Weiwei Sun and Lingyong Yan and Xinyu Ma and Shuaiqiang Wang and Pengjie Ren and Zhumin Chen and Dawei Yin and Zhaochun Ren},
  title     = {Is {ChatGPT} Good at Search? Investigating Large Language Models as Re-Ranking Agents},
  booktitle = {Proceedings of the 2023 Conference on Empirical Methods in Natural Language Processing},
  pages     = {14918--14937},
  year      = {2023},
  doi       = {10.18653/v1/2023.emnlp-main.923}
}

@inproceedings{yoon2024listt5,
  author    = {Soyoung Yoon and Eunbi Choi and Jiyeon Kim and Hyeongu Yun and Yireun Kim and Seung-won Hwang},
  title     = {{ListT5}: Listwise Reranking with Fusion-in-Decoder Improves Zero-shot Retrieval},
  booktitle = {Proceedings of the 62nd Annual Meeting of the Association for Computational Linguistics},
  pages     = {2287--2308},
  year      = {2024},
  doi       = {10.18653/v1/2024.acl-long.125}
}

@inproceedings{reddy2024first,
  author    = {Revanth Gangi Reddy and JaeHyeok Doo and Yifei Xu and Md Arafat Sultan and Deevya Swain and Avirup Sil and Heng Ji},
  title     = {{FIRST}: Faster Improved Listwise Reranking with Single Token Decoding},
  booktitle = {Proceedings of the 2024 Conference on Empirical Methods in Natural Language Processing},
  pages     = {8642--8652},
  year      = {2024},
  doi       = {10.18653/v1/2024.emnlp-main.491}
}

@inproceedings{ong2025routellm,
  author    = {Isaac Ong and Amjad Almahairi and Vincent Wu and Wei-Lin Chiang and Tianhao Wu and Joseph E. Gonzalez and M. Waleed Kadous and Ion Stoica},
  title     = {{RouteLLM}: Learning to Route {LLM}s with Preference Data},
  booktitle = {Proceedings of the 13th International Conference on Learning Representations (ICLR)},
  year      = {2025}
}

@inproceedings{song2020mpnet,
  author    = {Kaitao Song and Xu Tan and Tao Qin and Jianfeng Lu and Tie-Yan Liu},
  title     = {MPNet: Masked and Permuted Pre-training for Language Understanding},
  booktitle = {Advances in Neural Information Processing Systems},
  volume    = {33},
  pages     = {16857--16867},
  year      = {2020}
}

@article{reimers2019sentencebert,
  author  = {Nils Reimers and Iryna Gurevych},
  title   = {Sentence-BERT: Sentence Embeddings using Siamese BERT-Networks},
  journal = {Proceedings of EMNLP-IJCNLP},
  pages   = {3982--3992},
  year    = {2019}
}

@misc{cai2026openjev,
  author       = {Zefan Cai},
  title        = {Open-Jev-9B},
  year         = {2026},
  howpublished = {Hugging Face model repository},
  url          = {https://huggingface.co/ZefanCai/Open-Jev-9B},
  note         = {Accessed 2026-09-26}
}

@misc{qwen2026qwen35,
  author       = {{Qwen Team}},
  title        = {Qwen3.5-9B},
  year         = {2026},
  howpublished = {Hugging Face model repository},
  url          = {https://huggingface.co/Qwen/Qwen3.5-9B},
  note         = {Accessed 2026-09-26}
}

@article{nasika2026jina,
  author  = {Christina Nasika and Feng Wang and Antonis Minas Krasakis and Han Xiao},
  title   = {jina-reranker-v3.5: An Efficient Listwise Reranker with Hybrid Attention and Self-Distillation},
  journal = {arXiv preprint arXiv:2607.18152},
  year    = {2026}
}

@article{li2025fadti,
  author  = {Runze Li and Hanchen Wang and Wenjie Zhang and Binghao Li and Yu Zhang and Xuemin Lin and Ying Zhang},
  title   = {{FADTI}: Fourier and Attention Driven Diffusion for Multivariate Time Series Imputation},
  journal = {arXiv preprint arXiv:2512.15116},
  year    = {2025},
  doi     = {10.48550/arXiv.2512.15116}
}

@inproceedings{cao2007listnet,
  author    = {Zhe Cao and Tao Qin and Tie-Yan Liu and Ming-Feng Tsai and Hang Li},
  title     = {Learning to Rank: From Pairwise Approach to Listwise Approach},
  booktitle = {Proceedings of the 24th International Conference on Machine Learning (ICML)},
  pages     = {129--136},
  year      = {2007},
  doi       = {10.1145/1273496.1273513}
}

@inproceedings{xia2008listmle,
  author    = {Fen Xia and Tie-Yan Liu and Jue Wang and Wensheng Zhang and Hang Li},
  title     = {Listwise Approach to Learning to Rank: Theory and Algorithm},
  booktitle = {Proceedings of the 25th International Conference on Machine Learning (ICML)},
  pages     = {1192--1199},
  year      = {2008},
  doi       = {10.1145/1390156.1390306}
}

@inproceedings{schuster2022calm,
  author    = {Tal Schuster and Adam Fisch and Jai Prakash Gupta and Mostafa Dehghani and Dara Bahri and Vinh Q. Tran and Yi Tay and Donald Metzler},
  title     = {Confident Adaptive Language Modeling},
  booktitle = {Advances in Neural Information Processing Systems},
  volume    = {35},
  year      = {2022}
}

@inproceedings{gupta2024cascades,
  author    = {Neha Gupta and Harikrishna Narasimhan and Wittawat Jitkrittum and Ankit Singh Rawat and Aditya Krishna Menon and Sanjiv Kumar},
  title     = {Language Model Cascades: Token-Level Uncertainty and Beyond},
  booktitle = {International Conference on Learning Representations (ICLR)},
  year      = {2024}
}

@inproceedings{dekoninck2025routing,
  author    = {Jasper Dekoninck and Maximilian Baader and Martin Vechev},
  title     = {A Unified Approach to Routing and Cascading for {LLM}s},
  booktitle = {Proceedings of the 42nd International Conference on Machine Learning (ICML)},
  series    = {Proceedings of Machine Learning Research},
  volume    = {267},
  pages     = {12987--13010},
  year      = {2025}
}

@inproceedings{zhang2024jellyfish,
  author    = {Haochen Zhang and Yuyang Dong and Chuan Xiao and Masafumi Oyamada},
  title     = {Jellyfish: Instruction-Tuning Local Large Language Models for Data Preprocessing},
  booktitle = {Proceedings of the 2024 Conference on Empirical Methods in Natural Language Processing (EMNLP)},
  pages     = {8754--8782},
  year      = {2024},
  doi       = {10.18653/v1/2024.emnlp-main.497}
}

@inproceedings{seedat2024matchmaker,
  author    = {Nabeel Seedat and Mihaela van der Schaar},
  title     = {Matchmaker: Self-Improving Large Language Model Programs for Schema Matching},
  booktitle = {Generative AI and Health Workshop at NeurIPS},
  year      = {2024},
  url       = {https://arxiv.org/abs/2410.24105}
}

@article{sheetrit2024rematch,
  author  = {Eitam Sheetrit and Menachem Brief and Moshik Mishaeli and Oren Elisha},
  title   = {{ReMatch}: Retrieval Enhanced Schema Matching with {LLM}s},
  journal = {arXiv preprint arXiv:2403.01567},
  year    = {2024},
  doi     = {10.48550/arXiv.2403.01567}
}

@article{chen2026construm,
  author  = {Houming Chen and Zhe Zhang and H. V. Jagadish},
  title   = {{ConStruM}: A Structure-Guided {LLM} Framework for Context-Aware Schema Matching},
  journal = {arXiv preprint arXiv:2601.20482},
  year    = {2026},
  doi     = {10.48550/arXiv.2601.20482}
}

\end{document}